# Imaging Dissipation and Hot Spots in Carbon Nanotube Network Transistors

David Estrada,[1] and Eric Pop[1, 2, *]

[1]*Dept. of Electrical and Computer Engineering, Micro and Nanotechnology Laboratory, University of Illinois, Urbana-Champaign, IL 61801, USA*

[2]*Beckman Institute, University of Illinois, Urbana-Champaign, IL 61801, USA*

We use infrared thermometry of carbon nanotube network (CNN) transistors and find the formation of distinct hot spots during operation. However, the *average* CNN temperature at breakdown is significantly lower than expected from the breakdown of individual nanotubes, suggesting extremely high regions of power dissipation at the CNN junctions. Statistical analysis and comparison with a thermal model allow the estimate of an upper limit of the average tube-tube junction thermal resistance, $\sim 4.4 \times 10^{11}$ K/W (thermal conductance ~2.27 pW/K). These results indicate that nanotube junctions have a much greater impact on CNN transport, dissipation, and reliability than extrinsic factors such as low substrate thermal conductivity.

[*]Contact: epop@illinois.edu

Random networks of single-wall carbon nanotubes (CNTs) are of interest for integrated circuits and display drivers[1] on flexible or transparent substrates, particularly where they could exceed the performance of organic or amorphous thin-film transistors (TFTs). A common problem of such TFTs is that they are often placed on low thermal conductivity substrates like glass or plastics, leading to self-heating effects and reduced reliability,[2] topics not yet explored in carbon nanotube network (CNN) transistors. An additional concern with CNNs is that performance and reliability may be limited by high electrical[3] and thermal[4-7] inter-tube junction resistances. For CNNs this could result in large temperature increases (hot spots) at the CNT junctions, which greatly exceed the average temperature of the device channel.

In this study, we use infrared (IR) thermal imaging[8] and electrical breakdown thermometry[9] to investigate power dissipation in CNNs. We show that under high bias stress, devices fail with a minimal rise in *average* temperature. Furthermore, we show power dissipation can be localized at so-called "hot spots" in the CNN, which can be detrimental to TFT applications. In addition, we introduce a model to extract the average thermal resistance between CNNs and the substrate ($R_C$), as well as the CNT junction thermal resistance ($R_J$). Our results indicate that the latter is the key limiting factor in CNN performance, dissipation and reliability.

The CNN devices in this work are typically networks of single-wall CNTs fabricated on $SiO_2$(90 nm)/Si substrates, as outlined in the supplementary information.[10] All IR thermometry measurements are performed at a background temperature $T_0 = 70$ °C for optimum IR microscope sensitivity.[8] The highly n-doped Si acts as a back gate, set to $V_G < $ -15 V here, such that both metallic and semiconducting CNTs are turned "on." We acquire IR images at increasing source-drain bias ($V_{SD}$) and, surprisingly, we find the imaged channel temperature increases very little, even near the device breakdown. For instance, the maximum temperature rise imaged[10] in the high density (HD)[11] CNN shown in Figs. 1(b) and 1(c) is $\Delta T \approx 108$ °C at a power $P = I_D V_{SD} = $ 25 mW. Moreover, the temperature in the channel is non-uniform, with distinct hot spots which depend on the local CNN density variations and the CNT percolative pathways.

Lower density (LD) CNNs [Fig. 2(a)] do not provide as strong an IR thermal signal,[10] but facilitate analysis as the number of CNT junctions can be readily examined and counted by SEM,[11] as will be shown below. The measured power vs. voltage of LD and HD[11] CNNs up to breakdown (BD) are shown in Fig. 2(b). For both we note a sharp and irreversible drop, corres-ponding to $P_{BD}$ ~ 6.7 and 30 mW for the LD and HD devices, respectively. This signals a catastrophic break of the CNN, also noted when the LD device cannot be recovered on a subsequent sweep [dashed line in Fig. 2(b)]. In addition, we note the breakdown location of the film from Fig. 2(c) bears the imprint of the hot spot formation in the overlaid image of Fig. 2(d).

We now focus on the LD device to understand how measured $P_{BD}$ corresponds to $T_{BD}$ and the temperature measured by IR microscopy. In general, the power and temperature rise of a device are related through its thermal resistance,[12] here $T_{BD} - T_0 = P_{BD} \cdot R_{TH}$ at breakdown. We develop a thermal resistance model as shown in Fig. 3(a), and we assume the well-known $T_{BD} = 600\ °C$ for CNTs in air,[9] recalling that $T_0 = 70\ °C$. To simplify the analysis we assume uniform power dissipation across the CNN, although we know this is not strictly the case due to the percolative transport, as well as the imaged temperature profile [Fig. 2(d)]. However, as we will show, this allows us to determine a quantitative upper bound on the CNT junction resistance, $R_J$.

We note that power is dissipated both at the CNT junctions, and along the length of the CNTs in contact with the $SiO_2$.[13] This requires knowledge of the junction area fill factor ($\gamma_J$) with respect to the CNN area ($A_C$). To determine $\gamma_J$ we first extract the area fill factor of the network ($\gamma_C$) by analyzing SEM images. The images are imported to a matrix form in Matlab[14] and a threshold contrast is chosen to designate areas occupied by CNTs,[10] as shown in Fig. 3(b). The proportion of matrix elements with values above threshold is ~0.72, which is a significant overestimate of the true areal coverage ($\gamma_C$) as CNT diameters appear much larger under SEM, $30 < \langle d' \rangle < 80$ nm. Choosing $\langle d' \rangle \approx 50$ nm we estimate the total length of CNTs in the network, $L_C \approx 7.2$ mm, from $\gamma_C = \langle d' \rangle \sum L_C / A$, where the device area is $A = W \cdot L$. The actual area of the CNN is $A_C \approx d \cdot L_C \approx 14.4\ \mu m^2$, with a true device area fill factor $\gamma_C \approx 0.03$, where $d \approx 2$ nm is the real CNT diameter averaged from atomic force microscopy (AFM) analysis. (We return to the effect of variability introduced by $\langle d' \rangle$ from SEM analysis after extracting $R_J$ below.)

We estimate the total CNT-CNT junction area as $A_{JTOT} \approx A_J \cdot (n_J\ A)$, where $A_J$ is the average area of a CNT junction and $n_J$ is the junction density per device area $A$. We note the junction area depends on the angle of intersection ($\theta$) of CNTs in the random network, i.e. $A_J = d^2/\sin(\theta)$. Here we use image analysis software[14] to determine average values for $n_J$, $A_J$, and $\theta$, as shown by histograms in Fig. 3(c). We find $A_J = 4.69 \pm 0.93\ nm^2$, $\theta = 98 \pm 28°$, and $n_J \approx 26\ \mu m^{-2}$. Thus, the density of junctions in the network $\gamma_J = A_{JTOT}/A_C = 0.0042$, which completes the inputs needed

4for the thermal model in Fig. 3(a). We note that in general[3] $n_J$ will be proportional to CNN density and inversely with CNT segment lengths[13] between junctions. Therefore, when modeling other devices, it is important to carefully estimate $n_J$ for the particular CNN.

To find the total thermal resistance[12] of the CNN, we include the Si substrate thermal resistance $R_{Si} = 1/(2\kappa_{Si} A^{1/2})$, the SiO$_2$ thermal resistance $R_{ox} = t_{ox}/(\kappa_{ox} A_C)$, and the CNT-SiO$_2$ thermal boundary resistance of the network $R_C = 1/(gL_C)$. Here $t_{ox} = 90$ nm, $\kappa_{ox} \approx 1.4$ W m$^{-1}$ K$^{-1}$, $\kappa_{Si} \approx 100$ W m$^{-1}$ K$^{-1}$, and $g \approx 0.3$ W K$^{-1}$ m$^{-1}$ for CNTs of diameter ~2 nm near breakdown.[9] This gives $R_{Si} = 223.6$ K W$^{-1}$, $R_{ox} = 4.46 \times 10^3$ K W$^{-1}$, and $R_C = 462.9$ K W$^{-1}$, respectively.

We can now calculate the temperature rise at the SiO$_2$-Si interface, $\Delta T_{Si} = T_{Si} - T_0 = P_{BD} R_{Si} \approx 1.5$ K. This is a good match with the temperature measured by the IR imaging system for this device, considering that most IR signal originates from the top of the heated Si substrate.[8,10] The temperature drop across the SiO$_2$ is $\Delta T_{ox} = T_{ox} - T_{Si} = P_{BD} R_{ox} \approx 29.9$ K, and the temperature drop across the CNT-SiO$_2$ interface is[15] $\Delta T_C = T_C - T_{ox} = (1 - \gamma_J/2) P_{BD} R_C \approx P_{BD} R_C = 3.1$ K. Thus, the average temperature of the CNN *without* considering the effect of the junctions is merely $T_C \approx 104.5$ °C, much smaller than the breakdown temperature of CNTs in air, $T_{BD} \approx 600$ °C. This remains the case even when variability of the CNT-SiO$_2$ thermal coupling[9] ($g$) and that of the apparent diameter in SEM $\langle d' \rangle$ are taken into account. In other words, considering $g = 0.3 \pm 0.2$ W K$^{-1}$ m$^{-1}$ and $30 < \langle d' \rangle < 80$ nm in our analysis leads to a range $T_C \approx 90$–135 °C.

We suggest that the "missing" temperature difference is due to highly localized hot spots associated with the CNT junctions, which cannot be directly visualized by the IR thermometry. This is consistent with the emerging picture of CNT junctions being points of high electrical[3] and thermal[4-7] resistance. Consequently, we can extract the thermal resistance due to all CNT junctions ($R_{JTOT}$) in the network acting in parallel:[15]

$$R_{JTOT} = \frac{T_{BD} - T_C}{\frac{1}{2}\gamma_J P_{BD}} = \frac{T_{BD} - \Delta T_C - \Delta T_{ox} - \Delta T_{Si} - T_0}{\frac{1}{2}\gamma_J P_{BD}} \quad (1)$$

which is bound between 2.1–5.9 × 10$^7$ K W$^{-1}$ when allowing for uncertainty in $g$ and $\langle d' \rangle$ as above. $R_{JTOT}$ is several orders of magnitude greater than any other thermal resistance in the network, and remains dominant even if the SiO$_2$ were replaced with a substrate ten times more thermally insulating (e.g. plastics). If substrates with much higher thermal conductivity than SiO$_2$ are used (e.g. sapphire) the CNN junction thermal resistance is even more of a limiting factor.

We now estimate the thermal resistance of a single CNT junction as $R_J \approx R_{JTOT} \cdot (n_J A) \approx 4.4 \times 10^{11}$ K W$^{-1}$, equivalent to a thermal conductance $G_J \approx 2.27$ pW K$^{-1}$. We note it is likely that not all counted CNT junctions conduct current despite our effort to deliberately gate (turn on) the semiconducting CNTs. Thus, our estimate of CNT junction thermal resistance (conductance) represents an upper (lower) limit. Furthermore, accounting for the variability in CNT-SiO$_2$ coupling and $\langle d' \rangle$ from SEM analysis, we can place bounds on our estimate, $R_J \approx 2.7\text{–}7.6 \times 10^{11}$ K W$^{-1}$ ($G_J \approx 1.3\text{–}3.6$ pW K$^{-1}$). The $R_J$ obtained here is in good agreement with experimental results for bulk *single-wall* CNTs,[6] $\sim 3.3 \times 10^{11}$ K W$^{-1}$, and one order magnitude greater than measurements of intersecting *multi-wall* CNTs,[7] as would be expected. Our average CNT junction thermal resistance normalized by the average contact area from Fig. 3(c), is $r_J \approx 2.1 \times 10^{-6}$ m$^2$ K W$^{-1}$. This is one order of magnitude greater than $\sim 10^{-7}$ m$^2$ K W$^{-1}$ predicted by molecular dynamics simulations (MD) for overlapping (10,10) CNTs with 3.4 Å separation,[4,6] perhaps due to idealized conditions in the simulation or imperfection in the experiments.

To further understand the large apparent thermal resistance at CNT junctions, we point out that this is not only a function of the small overlap area $A_J$, but also of the average CNT separation and van der Waals (vdW) interaction.[4,6] Under the harmonic approximation, the spring constant between pairs of atoms is $K = 72\varepsilon/(2^{1/3}\sigma^2)$ from a simplified Lennard-Jones (LJ) 6-12 potential,[16] where $\varepsilon$ is related to the depth of the potential well, and $\sigma$ is a length parameter. Using parameters from Refs. [9,17] we find $K_{C\text{-}C} < K_{C\text{-}ox}/2$, i.e., the CNT-CNT thermal coupling is weaker than the CNT-SiO$_2$ thermal coupling per pair of atoms. This simple analysis does not account for the exact shape of the CNTs[9,17] or the role of SiO$_2$ surface roughness,[9] and thus further work must consider these effects to investigate the relatively "high" experimentally observed thermal resistance at single-wall CNT junctions.

In conclusion, we have directly imaged power dissipation in CNN transistors using IR microscopy. We found local hot spots in power dissipation detected by IR correlate to the subsequent breakdown of the network mapped by SEM. Nevertheless, these hot spots do not account for the CNN breakdown at relatively low *average* temperatures, <180 °C. Instead, our analysis suggests the CNN breakdown occurs at the highly resistive CNT-CNT junctions, allowing us to extract the junction thermal resistance $R_J \approx 4.4 \times 10^{11}$ K W$^{-1}$ (conductance 2.27 pW K$^{-1}$). These findings suggest that transport, dissipation, and reliability of CNN devices is limited by the CNT junctions rather than extrinsic factors such as low substrate thermal conductivity.

We acknowledge support from NSF grant CAREER ECCS 0954423, the NRI through the Nano-CEMMS Center, the Micron Technology Foundation, and the NDSEG Fellowship. We are indebted to A. Liao, J.D. Wood, and Z.-Y. Ong for helpful discussions and technical support.

**FIGURES**

**Two-column positioning:**

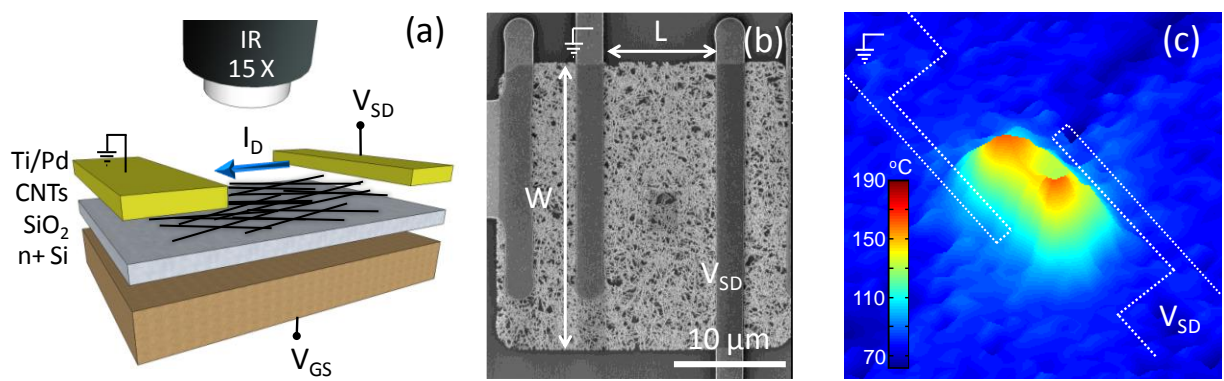

**FIG. 1:** (a) Schematic of CNN device and experimental setup. (b) Scanning electron microscopy (SEM) of HD CNN ($W/L \approx 25/10$ μm) before IR imaging and CNT breakdown. (c) Temperature of device in (b) measured at $P \approx 25$ mW, in air, with $T_0 = 70$ °C. The non-uniform temperature profile is indicative of percolative transport in such CNNs.

**Single-column positioning:**

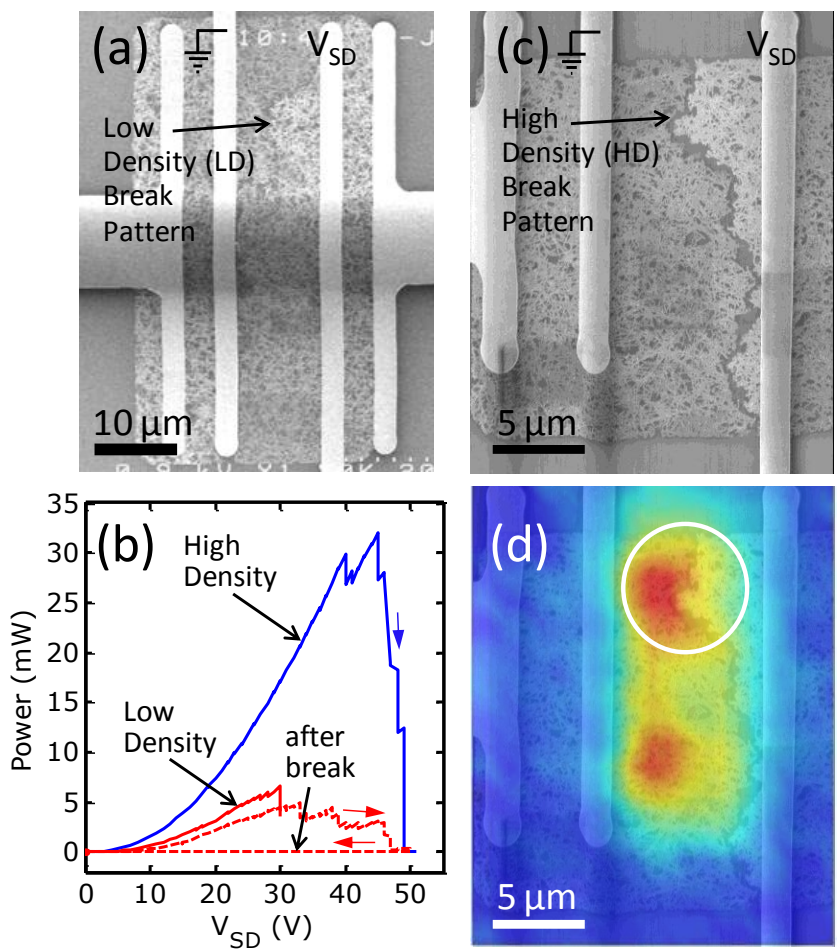

**FIG. 2:** (a) SEM image of LD device ($W/L \approx 50/10$ μm). (b) Measured power vs. voltage up to breakdown of LD device from (a) and HD device from (c). In both cases, large drops in power mark breaking of the CNN. The dashed line shows a second sweep of the LD device, taken after the first test was stopped at the $V_{SD} = 30$ V break. Small arrows indicate sweep directions. (c) SEM image of HD device from Fig. 1(b) after breakdown. (d) Measured temperature just before breakdown, at $P = 25$ mW from Fig. 1(c), overlaid onto the SEM from (c). The circled breakdown location bears the imprint of the adjacent hot spot. Although the breakdown occurs too fast to be imaged by the IR camera, we suspect the initial CNN break occurred at the upper hot spot, leading to a rerouting of the current pathways to cause the subsequent full break.

**Two-column positioning:**

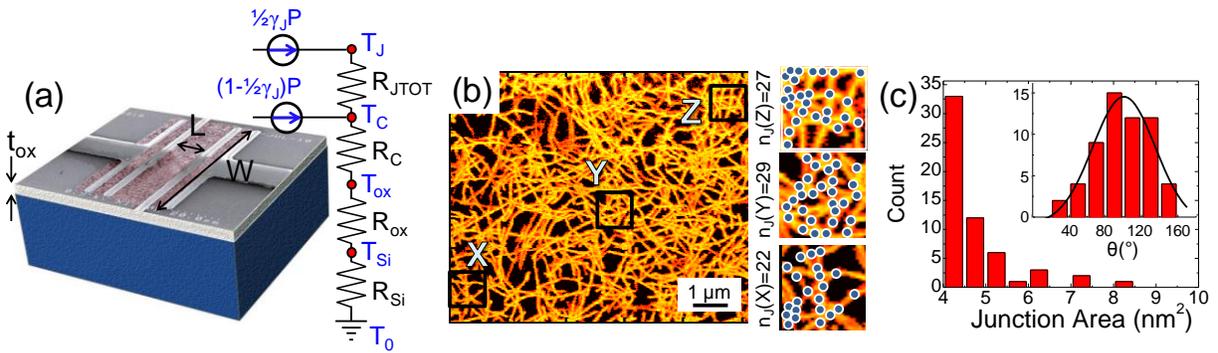

**FIG. 3:** (a) Thermal resistance model used to evaluate CNN dissipation and estimate the temperature differences, including from CNT junctions. (b) Processed SEM image of part of the LD device [from Fig. 2(a)] used for analysis of the total CNN length ($L_C$), area ($A_C$), and junction density ($n_J$). Highlighted portions of the SEM are magnified and the number of CNT junctions (dots) are counted to obtain averages. (c) Histogram of average CNT junction area $A_J$ and (inset) angle of intersection $\theta$.

**Supporting Online Materials** for "Imaging Dissipation and Hot Spots in Carbon Nanotube Network Transistors" by D. Estrada, and E. Pop, University of Illinois, Urbana-Champaign, U.S.A. (2011)

**1. Carbon nanotube network (CNN) device fabrication:** CNN devices used in this study were grown using an Etamota chemical vapor deposition (CVD) system. Low density devices were fabricated with ferritin catalyst following [SR-1]. High density devices were made by depositing ~2 Å Fe catalyst by e-beam evaporation. In both cases the catalysts were placed onto 90 nm $SiO_2$ on highly n-doped Si which acts as a back gate. Substrates were annealed at 900 °C in an Ar environment, followed by CNT growth for 15 minutes under $CH_4$ and $H_2$ flow. Standard photolithographic techniques were used to pattern the CNN by oxygen plasma etching, and the electrodes (Ti/Pd 1/40 nm) by lift-off, as shown in Fig. 1. Electrical and thermal measurements were performed using a Keithley 2612 dual channel source-meter and a QFI InfraScope II infrared (IR) microscope, respectively.

**2. Infrared Measurement Technique:** Before performing IR measurements of the CNN-TFTs, we acquire a reference radiance image which is used to calculate the emissivity at each detector pixel. This is done without biasing the device, at a background temperature $T_0 \sim 70$ °C for optimum IR microscope sensitivity [SR-2]. We then measure the background temperature with the IR scope to confirm the setup, verifying all pixels measure $T_0$.

**3. Infrared Properties of $SiO_2$ and Real Temperature of CNT junctions:** We can assume the $SiO_2$ is effectively transparent for near-IR radiation, because the thickness of the $SiO_2$ layer (90 nm) is much less than the optical depth for $SiO_2$ at these wavelengths. The optical depth for highly doped Si is much smaller and the temperature in the Si is highest near the Si-$SiO_2$ interface [SR-2; SR-3]. Hence, the IR Scope is effectively reading a thermal signal corresponding to a combination of the CNN temperature and that of the Si substrate near the Si-$SiO_2$ interface [SR-3].

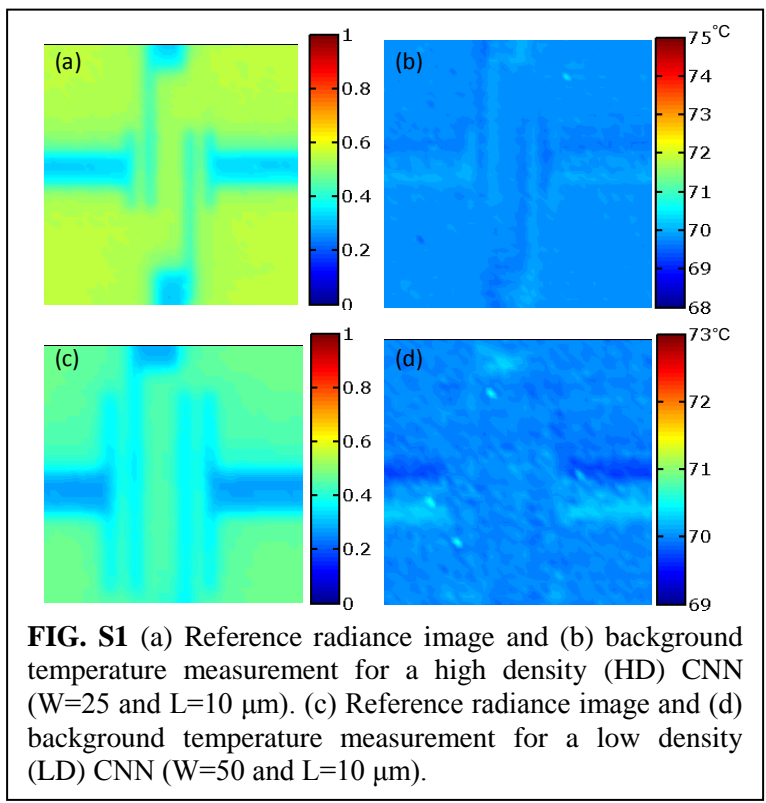

**FIG. S1** (a) Reference radiance image and (b) background temperature measurement for a high density (HD) CNN (W=25 and L=10 μm). (c) Reference radiance image and (d) background temperature measurement for a low density (LD) CNN (W=50 and L=10 μm).

To estimate the average temperature of the CNN given the temperature reported by the IR scope, we follow [SR-2] and the model in Fig. 3(a) in our main text. Thus, $(T_C-T_0) = (T_{Si}-T_0)(R_C+R_{ox}+R_{Si})/R_{Si}$.

Similarly, we can estimate the ratio between the T rise of the CNT junctions in the LD device and that of the Si surface as $(T_J-T_0)/(T_{Si}-T_0) = \frac{1}{2} \gamma_J(R_{JTOT}+R_C+R_{ox})/R_{Si} \approx 326$.

This agrees well with the imaged T profile of the LD device in Fig. S2. Here, the imaged temperature rise is only 1.5 °C. The actual temperature of the junctions near the breakdown power is nearly ~560 °C, consistent with the breakdown temperature of CNTs in air (see main text).

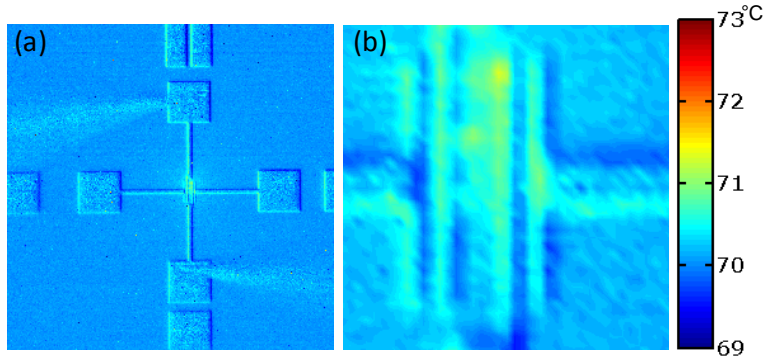

FIG. S2 (a,b) Temperature profile of the low-density (LD) device in Fig. S1(c-d) taken at a power of ≈ 5 mW and a background temperature $T_0 = 70$ °C. The non-uniform temperature profile is indicative of percolative transport in CNT devices.

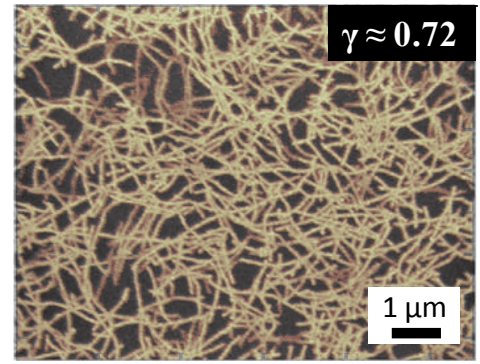

**FIG. S3** (a) Overlay of raw SEM data from Fig. 3(c) and Matlab modified SEM image, as used for analysis of the CNN length ($L_C$), area ($A_C$), and junction density ($n_J$). The apparent CNN area fill factor is ~0.72, which is an over-estimate due to the large apparent CNT diameter under SEM. The actual area fill factor for this network was closer to $\gamma_C = 0.03$ (see main text).

**4. Temperature Estimate of HD CNN:** While the temperature estimates of the LD CNN are given with comprehensive detail in the main text, this is not immediately possible for the HD CNN because the number of junctions and CNTs are not as easily countable. Nevertheless, to obtain the true temperature rise in Figs. 1(c) and 2(d) we perform the following estimate. Since the current of the HD device is ~5x that of the LD device, but their electrode separation is the same (10 μm), we surmise that $L_{C,HD} \sim 5 L_{C,LD} \sim 36$ mm. On the other hand, we note that the area of the HD device, $A_{HD} \sim A_{LD}/2 \sim 250$ μm$^2$. Thus, from the (more exact) LD device thermal resistances obtained in the main text, we estimate the same for the HD device as: $R_{C,HD} \sim R_{C,LD}/5 \sim$ 93 K/W, $R_{ox,HD} \sim R_{ox,LD}/5 \sim 892$ K/W and $R_{Si,HD} \sim \sqrt{2} R_{Si,LD} \sim 317$ K/W.

From these, we obtain the ratio between the T rise of the HD CNN vs. that imaged by IR is $(T_C - T_0) = (T_{Si} - T_0)(R_C + R_{ox} + R_{Si})/R_{Si} \sim 4.1$. Thus, since the peak T rise measured by IR for the HD CNN is ~ 26.3 K, the true peak temperature rise of the HD CNN is $\Delta T_{HD} \sim 108$ K (main text, page 1), or a maximum temperature $T_{HD} \sim 70 + 108 \sim 178$ °C [main text, Figs. 1(c) and 2(d)].